\documentclass[prb,preprint]{revtex4-1} 
\usepackage{graphicx} 

\newcommand{\be}{\begin{equation}}
\newcommand{\ee}{\end{equation}}
\newcommand{\ba}{\begin{eqnarray}}
\newcommand{\ea}{\end{eqnarray}}
\newcommand{\nn}{\nonumber}
\newcommand{\la}{\label} 
\newcommand{\e}{{\rm e}}

\begin{document}

\title{A truly elementary proof of Bertrand's theorem}

\author{Siu A. Chin}
\affiliation{Department of Physics and Astronomy, Texas A\&M University,
College Station, TX 77843, USA}

\email{chin@physics.tamu.edu} 
\date{\today}

\begin{abstract}
An elementary proof of Bertrand's theorem is given by examining
the radial orbit equation, without needing to solve complicated
equations or integrals.
\end{abstract}

\maketitle 

\section{Introduction} 

Bertrand's theorem\cite{ber73} is a surprising conclusion of 
Newtonian mechanics, which states that \textit{only} the attractive linear and inverse-square 
forces can yield closed, if bounded, orbits. (For an English translation of Bertrand's
paper, see Santos, Soares, and Tort;\cite{san11} for a near-transcription of Bertrand's
proof using modern notation, see Greenberg.\cite{gre66}) 
Since the harmonic potential cannot be physically
extended over astronomical distances, Newton's inverse-square gravitational force 
is the only possible force that can yield closed and thereby orderly planetary orbits. 
Thus, in any universe in which Newtonian mechanics is valid over 
a wide range of energy scales, and if orderly planetary orbits are considered 
desirable, then there is no choice in choosing the law of gravitational attraction 
but that of an inverse-square force. For this reason, Bertrand's theorem has continued
to fascinate the imagination of all those who are aware of its implication.
 
The proof of this theorem has continued to attract attention, despite the
fact that it was introduced more than 140 years ago. It has been proved using 
global methods,\cite{ber73,arn78,qui96} perturbative expansions,\cite{bro78,gol81,zar02,ant06}  
inverse transforms,\cite{tik88,yve08,san09}
and by searching for additional constants of motion.\cite{mar92} 
All of these proofs require sophisticated techniques from
solving integral equations to doing complicated integrals. None can
be consider ``elementary" conceptionally. In this work, we provide a truly elementary
proof, arrived at by merely inspecting the orbital equation. 

As summarized by Grandati, B\'erard, and M\'enas,\cite{yve08} 
most proofs of Bertrand's theorem have three steps. 
First, identify potentials having a constant 
apsidal angle. Second, determine these potential's apsidal angles for
near-circular orbits. Third, show that only a linear restoring force and 
Newton's gravitational force can have constant apsidal angles that 
are rational multiples of $\pi$ for general non-circular orbits. Steps one and two are
essentially the same in most proofs, while diverse methods are used to demonstrate
step three.

In this work, we follow the original global approach of Bertrand.\cite{ber73}
Bertrand solved step three by a seemingly mathematical fiat---evaluating his
integral conveniently from zero to one for vanishing potentials and from one to 
zero for rising potentials. His integration limit of zero corresponds to setting 
one turning point of the orbit to infinity. Brown\cite{bro78} has therefore 
opined that the perturbative method of proof, which only considers finite orbits, should 
be preferred. However, Arnold,\cite{arn78} in his exercise-proof of Bertrand's theorem, 
justified Bertrand's choice of turning points as special limits in the energy. 
Since the required apsidal angle is independent of the energy, it can be computed at 
any convenient energy. How the energy determines the turning points in Arnold's proof 
has since been made very clear in the work of Castro-Quilant\'an, Del R\'io-Correa, 
and Medina.\cite{qui96} In this work, their proof is further simplified by eliminating 
the need to calculate any complicated integrals. The final result is an explication of Bertrand's
original proof in its simplest form. 

Up to now, nearly all proofs\cite{arn78,qui96,bro78,gol81,zar02,ant06,tik88,yve08,san09} 
of Bertrand's theorem, including Bertrand's own proof,\cite{ber73} 
are not simple because they are based on computing the apsidal angle directly.
In this work, an elementary proof is obtained by 
computing the apsidal angle {\it indirectly}. That is, instead of
trying to determine the inverse function $\theta(r)$, it is simpler to examine the 
orbital trajectory in its usual form of $r(\theta)$. 
This is the new insight of this work. 
  
For completeness, we review steps one and two in the following section, while always 
maintaining the orbital equation in a familiar form. 
Section~III contains our elementary demonstration of step three. 

\section{The orbital equations and the apsidal angle}

For a spherically symmetric radial potential $V(r)$, the constancy of 
energy $E$ and angular momentum $L$ gives
\ba
\frac{dr}{dt}&=&\pm\sqrt{\frac2{m}\left[ E-V_{\rm eff}(r) \right]},\\
\frac{d\theta}{dt}&=&\frac{L}{mr^2},
\ea
where $V_{\rm eff}(r)$ is the effective potential
\be
V_{\rm eff}(r)=\frac{L^2}{2 m r^2}+V(r).
\ee
We trade the independent-variable $t$ with that of $\theta$ to obtain
\be
\frac{dr}{d\theta}=\pm\frac{mr^2}{L}\sqrt{\frac2{m}\left[ E-V_{\rm eff}(r) \right]}.
\ee
Introducing $u=1/r$, with $du=-dr/r^2$, we then have
\be
\frac{du}{d\theta}=\mp\frac{m}{L}\sqrt{\frac2{m}\left[ E-V_{\rm eff}(r) \right]}.
\ee
Squaring both sides, we can rearrange this equation into the familiar 
conservation-of-energy form
\be
\frac12 m^*\left(\frac{du}{d\theta}\right)^2+V_{\rm eff}(u)=E,
\la{oreq}
\ee
where we have defined the effective mass
\be
m^*=\frac{L^2}{m},
\ee
and where the effective potential now reads
\be
V_{\rm eff}(u)=\frac12 m^*u^2+V\left(\frac1{u}\right).
\la{veff}
\ee
Equation (\ref{oreq}) is identical in form to a conventional one-dimensional
dynamics problem and is the only equation we need to prove Bertrand's theorem.
This equation implies that we should seek to determine the orbital trajectory as
$u(\theta)$, rather than the inverse function $\theta(u)$. 

In what follows, we will need to compare various forms of Eq.~(\ref{oreq}) to the case of the
harmonic oscillator:
\be
\frac12 m\left(\frac{dx}{dt}\right)^2+\frac12 k x^2=E,
\la{ho}
\ee
where the solution is well-known to be
\be
x(t)=A\cos(\omega t)\qquad{\rm with}\qquad \omega=\sqrt{\frac{k}{m}}.
\la{hosol}
\ee
(We only need to assume the simplest case of $x(0)=A$.) Thus, if Eq.~(\ref{oreq})
takes the form of Eq.~(\ref{ho}), then by identifying the corresponding
coefficients $m$ and $k$, one can immediate infer the corresponding angular
frequency $\omega$ and the solution using Eq.~(\ref{hosol}) with $t\rightarrow \theta$.

We begin by considering Eq.~(\ref{oreq}) for circular motion with constant $u_0=1/r_0$,
which occurs when
\be
V_{\rm eff}^\prime(u_0)=m^*u_0-u_0^{-2}\,V^\prime\left(\frac1{u_0}\right)=0.
\la{vp}
\ee
Small oscillations about $u_0$ can then be written as
\be
u(\theta)=u_0+\rho(\theta).
\ee
To determine $\rho(\theta)$, we expand $V_{\rm eff}$ about $u_0$ to second-order in $\rho$:
\be
V_{\rm eff}(u)=V_{\rm eff}(u_0)+\rho V_{\rm eff}^\prime(u_0)+\frac12\rho^2 V_{\rm eff}^{\prime\prime}(u_0)+\cdots .
\ee
Since the second term in this expansion vanishes by Eq.~(\ref{vp}) and the constant first term can be
absorbed into the potential energy, Eq.~(\ref{oreq}) takes the form
 \be
\frac12 m^*\left(\frac{d\rho}{d\theta}\right)^2+\frac12 V_{\rm eff}^{\prime\prime}(u_0)\rho^2=E^\prime
\la{rho},
\ee
where $E^\prime=E-V_{\rm eff}(u_0)$.
Comparing this to Eq.~(\ref{ho}), one can immediately identify $m=m^*$, $k=V_{\rm eff}^{\prime\prime}(u_0)$,
and the solution as
\be
u(\theta)=u_0+A\cos(\Omega\theta)
\la{uos}
\ee
with angular frequency
\be
\Omega=\sqrt\frac{V^{\prime\prime}_{\rm eff}(u_0)}{m^*}.
\la{omg}
\ee

Next, we proceed by using Eq.~(\ref{veff}) to write
\be
V^{\prime\prime}_{\rm eff}(u_0)=m^*
+2u^{-3}_0\,V^\prime\left(\frac1{u_0}\right)+u^{-4}_0\,V^{\prime\prime}\left(\frac1{u_0}\right),
\ee
and then using $m^*$ from Eq.~(\ref{vp}) to get
\ba
\Omega=
\sqrt\frac{
3V^\prime(r_0)
+r_0V^{\prime\prime}(r_0)}
{V^\prime(r_0)}
\la{omg2}.
\ea		 
Without loss of generality, we have assumed in Eq.~(\ref{uos}) that $u$ starts at 
$u_{\rm max}=u_0+A$ with $\theta=0$, and reaches $u_{\rm min}=u_0-A$ at the
apsidal angle $\theta=\theta_A$. For such a half-cycle oscillation, the argument of
the cosine function in Eq.~(\ref{uos}) changes by $\pi$. Hence, we have the near-circular
orbit result:
\be
{\Omega}\theta_A=\pi\quad\rightarrow\quad
\theta_A=\frac{\pi}{\Omega}.
\la{ap}
\ee

For a constant apsidal angle that is independent of $E$ and $L$, we
must have a constant $\Omega$. For stability of oscillations about a circular orbit, 
we must also have a real $\Omega>0$. Both constraints are satisfied if we set
\be
\frac{
3V^\prime(r)
+rV^{\prime\prime}(r)}
{V^\prime(r)} = c > 0.
\la{const}
\ee
Writing $W(r)=V^\prime(r)$ then gives
\ba
\frac1{W}\frac{dW}{dr}=\frac{(c-3)}{r},
\ea
and an elementary integration yields the solution
\be
W(r)=k r^{c-3}\quad{\longrightarrow}\quad V(r)=\frac{k}{c-2}r^{c-2},
\ee
where $k$ is an integration constant.  This form of the potential is a result of the fact that we are focusing on $r(\theta)$.
Had we tried to determine $\theta(r)$, we would have obtained the ``unnatural" looking
result of $W(r)=k r^{1/c-3}$, as in Bertrand's original work or in modern texts, such as Ref.~\onlinecite{ant06}.

Since $c>0$, it is convenient to set $\alpha=c-2$ so that
\be
V(r)=\frac{k}{\alpha}r^\alpha,\quad{\rm with}\quad\alpha>-2.
\la{vpot}
\ee
Thus, for a constant apsidal angle, the potential must be a {\it single} power-law
of the form given in Eq.~(\ref{vpot})---any linear combination of power laws will not yield a
constant $\Omega$ via Eq.~(\ref{omg2}). 

We note that in the limit $\alpha\rightarrow 0$, Eq.~(\ref{vpot})
correctly gives the logarithmic potential
\ba
V(r)&=&\lim_{\alpha\rightarrow 0}\frac{k}{\alpha}\e^{\alpha\ln(r)},\nn\\
&=&\lim_{\alpha\rightarrow 0}\frac{k}{\alpha}\left[1+\alpha\ln(r)+\frac12\alpha^2\ln^2(r)+\cdots \right], \nn\\
&=& k\ln(r),
\la{vlog}
\ea
since once again, a constant term in the potential can be ignored. 
Thus, there is no need to consider the logarithmic potential separately, 
if we use the form of the potential in Eq.~(\ref{vpot}).

Substituting Eq.~(\ref{const}) into Eq.~(\ref{omg2}) gives $\Omega=\sqrt{c}=\sqrt{2+\alpha}$,
and so the apsidal angle~(\ref{ap})
for perturbed oscillations about a circular orbit
has the explicit form
\be
\theta_A=\frac{\pi}{\sqrt{2+\alpha}}.
\la{apsid}
\ee
One then immediately sees that $\theta_A$ is a rational multiple of $\pi$ for
$\alpha=-1$ and $\alpha=2$.	All remaining values of $\alpha$ will be ruled out 
in the next section.

\section{Step three}

    Since the apsidal angle is constant for arbitrary $E$, one can 
determine $\theta_A$ for arbitrary orbits in any convenient limit of $E$.
As suggest by Arnold,\cite{arn78} a convenient limit for $\alpha>0$ is to let
$E \rightarrow \infty$. In this limit, Eqs.~(\ref{oreq}) and (\ref{veff}) allow us
to see that $u_{\rm max}$ is governed by
\be
\frac12 m^*u_{\rm max}^2=E.
\ee
Using this result, along with Eqs.~(\ref{veff}) and (\ref{vpot}), allows us to write
the scaled effective potential $v_{\rm eff}(u)=V_{\rm eff}(u)/E$ as
\be
v_{\rm eff}(x)=x^2+\frac{k}{\alpha}\left(\frac{m^*}{2}\right)^{\alpha/2}E^{-(2+\alpha)/2}x^{-\alpha},
\ee
where $x=u/u_{\rm max}$.
With increasing $E$, the second term on the right-hand-side becomes completely negligible except 
near $x\approx 0$, as illustrated in Fig.~1.  Thus, as $E\rightarrow\infty$,
the energy-scaled effective potential 
approaches a harmonic oscillator potential with an infinite ``wall'' at $x=0$
($u_{\rm min}\rightarrow0$).  In this limit, we have
\be
\frac12 m^*\left(\frac{du}{d\theta}\right)^2+\frac12 m^* u^2=E.
\la{halfh}
\ee
By comparing this to Eq.~(\ref{ho}), we can identify $m=m^*$, $k=m^*$, and hence
$\Omega=\sqrt{m^*/m^*}=1$.  But, the oscillation of $u$ from $u_{\rm max}$ to $u_{\rm min}=0$
is only a {\it quarter} of the harmonic cycle of Eq.~(\ref{halfh}), yielding
\be
\Omega\theta_A=\frac\pi{2}\quad{\rightarrow}\quad \theta_A=\frac\pi{2}.
\ee
Comparing this to Eq.~(\ref{apsid}), it is clear that only $\alpha=2$,
the harmonic oscillator potential, is
consistent in having the same constant apsidal angle at near-circular
orbit energy and as $E\rightarrow\infty$.

For $\alpha\rightarrow 0$, the potential is logarithmic with irrational
apsidal angle $\theta_A=\pi/\sqrt{2}$ and cannot yield any closed orbit.

For $\alpha<0$, we let $\alpha=-s$ with $0<s<2$ so that 
$V(r)=-(k/s)r^{-s}$. For these potentials, all bounded orbits have $E<0$, and a convenient
limit is to take $E \rightarrow 0$ (with $r_{\rm max} \rightarrow \infty$). 
In this case, Eq.~(\ref{oreq}) takes the form of
\be
\frac12 m^*\left(\frac{du}{d\theta}\right)^2+\frac12 m^* u^2-\frac{k}{s}u^s=0.
\ee
This equation will remain familiar if we have a positive constant,
resembling the energy, on the right-hand side. 
Hence, we divide every term by $u^s$, and move the negative term to the
right:
\be
\frac12 m^*u^{-s}\left(\frac{du}{d\theta}\right)^2+\frac12 m^* u^{2-s}=\frac{k}{s}.
\la{eq2}
\ee
Since the only equation we can solve by inspection is that of the harmonic oscillator, 
let's set $x^2=u^{2-s}$ so that
\be
(2-s)\frac{du}{u}=2\frac{dx}{x}.
\label{ux}
\ee
Now arrange Eq.~(\ref{eq2}) as
\be
\frac12 m^*u^{2-s}\left(\frac{1}{u}\frac{du}{d\theta}\right)^2+\frac12 m^* x^2 = \frac{k}{s},
\ee
and use Eq.~(\ref{ux}) to transform it into
\be
\frac12 m^*\left(\frac2{2-s}\right)^2\left(\frac{dx}{d\theta}\right)^2+\frac12 m^*x^2 = \frac{k}{s}.
\la{eq3}
\ee
Comparing this to Eq.~(\ref{ho}), we see that $m=m^*[2/(2-s)]^2$,
$k=m^*$, and therefore
\be
\Omega=\frac{2-s}2.
\ee
Since $r_{\rm max}=\infty$ implies that $x_{\rm min}=0$, the oscillation from
$x_{\rm max}$ to $x_{\rm min}=0$ is again only a quarter period and we have
\be
\Omega\theta_A=\frac\pi{2}\quad\rightarrow\quad\theta_A=\frac\pi{2-s}.
\ee
Comparing this to Eq.~(\ref{apsid}), the potential that can have
the same apsidal angle independent of energy must satisfy 
\be
\frac{\pi}{2-s}=\frac{\pi}{\sqrt{2-s}}.
\ee
This equation yields the unique solution $s=1$, corresponding to $\alpha=-1$ and
the inverse-square force law, thus completing our proof of Bertrand's theorem. 

\section{Conclusion}
In this work, we have provided a truly elementary proof of Bertrand's theorem.
This should help to make this important theorem 
of classical mechanics more accessible to undergraduate students.

\begin{acknowledgments}
This publication was made possible by NPRP GRANT \#5-674-1-114 
from the Qatar National Research Fund (a member of Qatar Foundation). The 
statements made herein are solely the responsibility of the author.
\end{acknowledgments}

\newpage
\begin{figure}
	\centerline{\includegraphics[width=8.5cm]{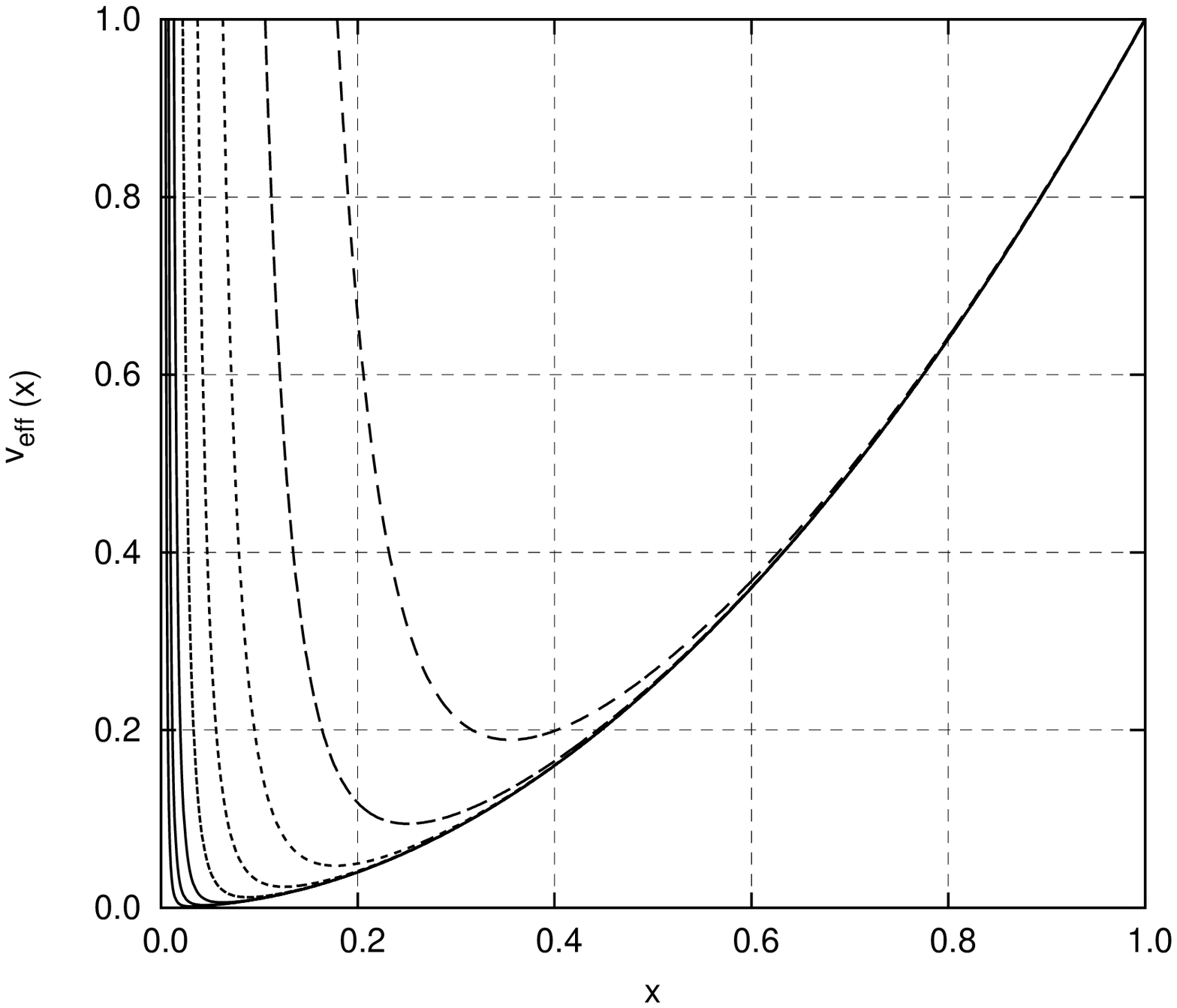}}
\caption{
Typical behaviors of the energy-scaled effective potential for rising potentials
at increasing $E$ (from dashed, to dotted, to solid lines).
Note that as $E\rightarrow\infty$, the (scaled) effective potential energy approaches
a harmonic oscillator potential with an infinite barrier at $x=0$.  
\label{fig1}}
\end{figure}

\end{document}